\documentclass{article}
\usepackage{lastpage}
\usepackage{graphicx}
\pdfoutput=1
\usepackage[T1]{fontenc}
\usepackage{float}
\usepackage{xcolor}
\usepackage{multirow}
\usepackage{diagbox}
\usepackage[toc,page,title,titletoc,header]{appendix}
\usepackage{listings}
\usepackage{pxfonts}
\usepackage{mathrsfs}
\usepackage[numbers,sort&compress]{natbib}
\usepackage{CJK}
\usepackage{theorem}
\usepackage[scale=0.75]{geometry}
 \pdfoutput=1
\newtheorem{thm}{Theorem}[section]

{
  
}

\usepackage{hyperref}
\usepackage{fancyhdr}
\begin{document}
 \pdfoutput=1

\date{}
\title{Research on ruin probability of risk model based on AR(1) series}
\maketitle

\begin{center} {Wenhao Li,\ Bolong Wang,\ Tianxiang Shen,\ Ronghua Zhu\ and~~Dehui Wang}
\end{center} \begin{center} {\emph{ (College of Mathematics,Jilin University,Changchun 130012)}} \end{center}
\newcommand{\upcite}[1]{\textsuperscript{\textsuperscript{\cite{#1}}}}
\bibliographystyle{plain}


%

{\textbf{Abstract}}:\ In this text, we establish the risk model based on AR(1) series and propose the basic model which has a dependent structure under intensity of claim number. Considering some properties of the risk model, we take advantage of newton iteration method to figure out the adjustment coefficient and estimate the exponential upper bound of ruin probability. This is significant to refine the research of ruin theory. As a result, our theory will help develop insurance industry stably.


{\textbf{Classification codes:\nocorr{O212.7}  \ \ \ \ \ \    Document id code:A}}\\



\section{Introduction}

The ruin probability was born in the last century, dating back to the doctoral dissertation which was proposed by a Sweden actuary in 1903. Basically, his work laid the foundation of our modern random model towards the non-life field. Nevertheless, the results made by Lundberg were not correspond to strict standard in Mathematics, as a result, statisticians, mathematicians and employees in financial industry, including many stake holders have been working on figuring out a more specific model which is much more suitable and precise in a modern circumstance throughout history. 
 With the guidance of classical risk model, people have achieved many precious results, from the theory made by Lundberg\upcite{Lundberg} to Cramer\upcite{Cramer}, who applied strict mathematical tool to figure out the initial value of $\theta$(0), which has been proved to be \ $\Phi(0)=\frac{1}{1+\theta}$. Also, the inequation $\phi(u)\le e^{-Ru}$ which corresponds to the Lundberg inequality and Lundberg-Cramer inequation is one of the achievements. Every coin has two sides and this is not an exception as well. Classical model like this is far too complex, ideal and unsuitable for the modern situation which is flexible and changeable. 
It¡¯s too numerous to mention one by one as modern risk theories spring up quickly. Wang\upcite{Wang D H} researched into the condition when premiums depend on claims and put forward the exponential upper bound of ruin probability. Fu\upcite{Fu K A} estimated the ruin probability under the condition of renewal risk model depending on time series. Meanwhile, Bao\upcite{5} proposed a risk model in specific field where is a Poisson process. Also, Joe\upcite{6} and Gu\upcite{7} researched into one model of applying the Copula function into the risk process of Erlang, and finally achieved the integral differential equation of the final ruin probability. Lee\upcite{8} takes advantage of several models in time series (AR(p) model, MA(q) model, ARMA(p, q) model and so on), analyzing ruin probability from three main areas: collecting quantities every unit time, claiming quantities and interests. Additionally, Chen\upcite{9} discussed the final ruin probability where compound-Poisson distribution is applied in constant interest , achieving the visible answer of final ruin probability where several settlement of claims are under the exponential distribution and made several discussion towards risk model with interest constraint. Chen\upcite{10} got the asymptotic property of final risk probability with interests where claims depend on each other and belong to the ERV group, but the interval of claims is i.i.d. Guo\upcite{11} was aimed at different financial environment and risk model with dependent structure, building up the asymptotic estimation or inequation of ruin probability towards initial surplus. Yu\upcite{12} used Vine Copula model, which is based on conditional Copula function and Vine graphical modeling tool to connect the edge distribution function of multivariate financial variables into a joint function. Mo\upcite{13} defined the Markov-dependent risk model rigorously and proved the equivalence theorem of this model. Zhang\upcite{14} assumed that the insurance and financial risks follow a wide type of dependence structure and the insurance risks are sub exponential. Wang\upcite{Wang Kaiyong} obtained the asymptotic estimates of the finite-time ruin probability of a dependent risk model with a constant interest rate by using the probability limiting theory and stochastic process. Wang\upcite{16} gave an asymptotically equivalent formula for the finite-time ruin probability of a nonstandard risk model with a constant interest rate. Cai\upcite{17} studied ruin probabilities in two generalized risk models. Generalized Lundberg inequalities for the ruin probabilities are derived by a renewal recursive technique and the rates of interest are assumed to have a dependent autoregressive structure. Badescu\upcite{18} constructed an analytically tractable fluid flow that leads to the analysis of various ruin-related quantities in the aforementioned risk model. 
 Chen\upcite{19} considered uniform asymptotics for the finite-time ruin probabilities of two kinds of nonstandard bidimensional renewal risk models with constant interest forces and diffusion generated by Brownian motions. Wang\upcite{20} studied the asymptotics of the finite-time ruin probability for a generalized renewal risk model with independent strong subexponential claim sizes and widely lower orthant dependent inter-occurrence times. Lefevre\upcite{21} aimed to present some useful methods that have been proposed so far for computing, or approximating, the probabilities of (non-) ruin of a company over any given horizon of finite length in the case of discrete claim severities. Gao\upcite{22} considered ruin problems in two generalized risk models. The effects of timing of payments and interest on the ruin problems in the models are studied. The rates of interest are assumed to have an autoregressive structure. Fu\upcite{23} considered a continuous-time renewal risk model, in which the claim sizes and inter-arrival times form a sequence of independent and identically distributed random pairs, with each pair obeying a dependence structure. Li\upcite{24} derived the asymptotic finite-time ruin probability and study the optimal allocation of the global initial reserve in the sense of minimizing the asymptotic ruin probability. 
  Yao\upcite{25} considered two discrete-time risk models, in which dependent structures of the payments and the interest force are considered. 
 Hansjorg Albrecher\upcite{26} considered a generalization of the classical ruin model to a dependent setting, where the distribution of the time between two claim occurrences depends on the previous claim size. Meng\upcite{27} constructed a model with a dependent setting where the time between two claim occurrences determines the distribution of the next claim size.

In the classical model,\ the number of claims, which is defined as N,\ is an integer. The quantitative analysis of current insurance industry is always based on the assumption that the claims form a sequence of independent identically distributed random variables, and they have nothing to do with time. However the assumption is not in the line of actual. In fact, a claim  in a specific insurance policy will be influenced by the previous claims to some extent, so there is a great dependence between them. Taking the Northeast China Region as an example, during the winter, there are lots of similar accidents occurring because of slippery roads.\ Meanwhile, the increase of the number of accidents will lead to the increase of the number of insurance in the next phase, and it will directly influence the claim of the next phase. Based on this, we introduce the claim intensity that has dependent structures, which is defined as $\Lambda_t$ in the risk model. Since the number of claims increases constantly over time, we define that the number of claims as $N_t$, obeying to the Poisson distribution with the intensity parameters, $\Lambda_{t}+t$. Because AR(1) series reflect dependency and stability, and it considers the influence of random factors and the close relationship with time. Therefore, we define that the intensity parameters,\ $\Lambda_{t} $, obeying to the AR(1), $\{N_t,\ t=1,2,\cdot \cdot \cdot \}$ and\ $\{X_i,i=1,2,\cdot \cdot \cdot \}$ are independent. This makes the claim of insurance companies more reasonable and effective. What's more, it promotes stable and healthy development of insurance industry in China. The following is the basic definition and derivation of the model.

We will make the number of claims(N(t)) for the stochastic process which will be changed over time. The number of claims of current phase will not only under the influence of various factors of current phase but also be influenced by the previous phase. Based on this, there is a strong dependence. Furthermore, the dependency of number of claims is reflected in the intensity of claim number. Therefore, we will introduce the intensity of claim number which has a high dependence, defined as\ $\Lambda_t$, to our model.

\begin{equation}\label{Lambda_{t}}
ln\Lambda_{t}=\alpha ln(\Lambda_{t-1}) + \varepsilonup_{t},\ t=1,2,\cdot \cdot \cdot ,\nonumber
\end{equation}
among them $\varepsilonup_{t}$ is white noise series, and
$$\varepsilonup_{t} \sim  N(\mu,\sigma^2),$$

namely \begin{eqnarray}\label{moxing}     
\left\{                        
\begin{array}{lll}       
S_t =\Sigma^{N_{t}}_{i=1}X_i, \\  
ln\Lambda_{t}=\alpha ln(\Lambda_{t-1}) + \varepsilonup_{t}, \\
\varepsilonup_{t} \sim  N(\mu,\sigma^2), \\
\end{array}              
\right.                       
\end{eqnarray}
which is the risk model we are about to establish.
Let \ $N_t$  be a Poisson counting process with parameter \ $\Lambda_t +t$ and it represents the number of claims in time period (0, t]. Parameter \ $\Lambda_t$ obeys AR(1) model. $\{N_t,t=1,2,\cdot \cdot \cdot \}$  and $\{X_i,i=1,2,\cdot \cdot \cdot \}$ are independent.
$X_i$\ denotes the i-th claim size and it obeys an exponent distribution with density\ $f(x)=\frac{1}{\theta}e^{-\frac{1}{\theta}x}, x>0, \theta >0$.


The above is the risk model we established based on AR(1) series.

\section{Properties of the Model and  Parameter Estimation}
\subsection{\emph{Properties of the Model}}
In order to estimate the parameters of the model by using the method of moment estimation, we first study the properties of the model parameters. Then we figure out the expectation, the variance and the third-order origin moment of the $S_t$ which represents the total claims.
First of all, we calculate the expectation of $S_t$.

Since\ $\{N_t, t=1,2,\cdot \cdot \cdot \}$\ and the amount of the claims $\{X_i, i=1, 2, \cdot \cdot \cdot \}$\ are independent, we have
\begin{eqnarray}\label{2}
E(S_t)&=&E[E(S_{t}\ |\   N_{t})]=E( N_{t})\ E(X_i) =E[E(N_{t}\ |\ \Lambda_{t})]\ E(X_i).\nonumber
\end{eqnarray}
For\ $N_{t}$\ obeys the poisson distribution with parameter \ $\Lambda_{t}+t$, we have\ $E(N_{t}\ |\ \Lambda_{t})=\Lambda_{t}+t$. Thus,
\begin{eqnarray}\label{2}
E(S_t)&=&E[\Lambda_{t}+t]\ E(X_i).\nonumber
\end{eqnarray}
Let $ln\Lambda_{t}=Y_t$, we have\ $\Lambda_{t}=e^{Y_t}$, where $ln\Lambda_{t}=Y_t$ obeys AR(1) which is  a wide stationary sequence, namely
\begin{equation}\label{Y_t}
Y_t=\alpha \ Y_{t-1}+\varepsilonup_{t},\ \alpha \in (-1,1),\ t=1,2,\cdot \cdot \cdot ,\nonumber
\end{equation}
According to the property of the wide stationary time series, we have
$$
E(Y_t)=\frac{\mu}{1-\alpha},Var(Y_t)=\frac{\sigma^2}{1-\alpha^2},\forall t=1,2,\cdot \cdot \cdot .
$$
Therefore,
$E[\Lambda_{t}]=E[e^{Y_t}]$\ , where $ Y_t \sim N(\frac{\mu}{1-\alpha},\frac{\sigma^2}{1-\alpha^2}), \forall t=1,2,\cdot \cdot \cdot $
Furthermore, we calculate the  mathematical expectation\ $E[\Lambda_{t}],$\ we get
$$
E[\Lambda_{t}]=E[e^{Y_t}]=e^{\mu/(1-\alpha)+\sigma^2/2(1-\alpha^2)}.\nonumber
$$
Finally we obtain the expectation of \ $S_t$\ \begin{equation}\label{E}
E(S_t)=E[\Lambda_{t}+t]\ E(X_i)=[e^{\mu/(1-\alpha)+\sigma^2/2(1-\alpha^2)}+t]E(X_i).
\end{equation}

Secondly, we calculate the  variance of\ $S_t$.
Note that\begin{eqnarray}\label{2}
Var(S_t)
      &=&E[ N_{t}]Var(X_i)+E^{2}(X_i)Var[N_{t}].\nonumber
\end{eqnarray}
Derived from the foregoing\ $E[ N_{t}]=E[E(N_{t}\ |\ \Lambda_{t})]=E[\Lambda_{t}+t]=e^{\mu/(1-\alpha)+\sigma^2/2(1-\alpha^2)}+t$, further we calculate the variance of\ $N_{t}$\ .
\begin{eqnarray}\label{2}
Var[N_{t}]=E[Var(N_{t}\ |\ \Lambda_{t})]+Var[E(N_{t}\ |\ \Lambda_{t})]=E[ \Lambda_{t}+t]+Var[\Lambda_{t}]=t+ E[ \Lambda_{t}]+Var[\Lambda_{t}],\nonumber
\end{eqnarray}
In this way, the calculation of the variance of the number of claims is transformed into the calculation of the expectation and the variance of \ $\Lambda_{t}$.
For
\begin{eqnarray}
Var[\Lambda_{t}]=E[\Lambda^{2}_{t}]-E^{2}[\Lambda_{t}],\nonumber
\end{eqnarray} we can further calculate second-order origin moment of\ $\Lambda_{t}$.
$$
E[\Lambda^{2}_{t}]=E[(e^{Y_{t}})^{2}]=E[e^{2Y_t}],
$$
In addition, $\varepsilonup_{t}$\ follows normal distribution (see (\ref{Y_t})), and the normal distribution is additive. We can get that\ $2Y_t$\ follows normal distribution,
$$2Y_t \sim N(\frac{2\mu}{1-\alpha},\frac{4\sigma^2}{1-\alpha^2}),$$
Therefore,
$$
E[\Lambda^{2}_{t}]=E[e^{2Y_t}]=e^{2\mu/(1-\alpha)+2\sigma^2/(1-\alpha^2)}.\nonumber
$$
so that the variance of \ $S_t$\ is
\begin{equation}\label{V}
Var(S_t)=[Var(X_i)+E^{2}(X_i)]\ [e^{\frac{\mu}{1-\alpha}+\frac{\sigma^2}{2(1-\alpha^2)}}+t]+E^{2}(X_i)[  e^{\frac{2\mu}{1-\alpha}+\frac{2\sigma^2}{1-\alpha^2}}- e^{\frac{2\mu}{1-\alpha}+\frac{\sigma^2}{1-\alpha^2}}].
\end{equation}

Finally we calculate the third-order moment of\ $S_t$. First of all, we derive the moment generating function of \ $S_{t}$. Note that
\begin{eqnarray}\label{St}
M_{S_t}(r)&=&E[e^{S_{t} \cdot r}]=E[e^{\Sigma_{i=1}^{N_t} X_{i} \cdot r}]=E[E(e^{\Sigma_{i=1}^{N_t} X_{i} \cdot r}|N_{t})]=E[\Pi_{i=1}^{N_t} E(e^{X_i \cdot r})]\nonumber\\
&=&E\{[M_x (r)]^{N_{t}}\}=E[e^{N_{t} \cdot lnM_x (r)}]=M_{N_t} [ln M_x (r)].
\end{eqnarray}
In this way,\ for the sake of the moment generating function of the \ $S_{t}$,\ we first derive the moment generating function of\ $N_{t}$,
\begin{eqnarray}\label{Nt}
M_{N_t}(r)&=&E[e^{N_{t} \cdot r}]=E[E(e^{N_{t} \cdot r}|\Lambda_{t})]=E[e^{(\Lambda_t +t)(e^r -1)}]\nonumber\\
&=&e^{t(e^r -1)}E[e^{\Lambda_t (e^r -1)}]=e^{t(e^r -1)} M_{\Lambda_t}(e^{r}-1).
\end{eqnarray}
In order to further figure out the moment generating function of\ $N_{t}$, secondly, we derive the moment generating function of\ $\Lambda_{t}$, that is
\begin{eqnarray}\label{Lambdat1}
M_{\Lambda_t}(r)=E[e^{r\Lambda_t}]=E[e^{re^{Y_t}}],
\end{eqnarray}
where\ $Y_t \sim N(\frac{\mu}{1-\alpha},\frac{\sigma^2}{1-\alpha^2})$.
We substitute (\ref{Lambdat1}) into (\ref{Nt}), it follows that

\begin{eqnarray}\label{Nt1}
M_{N_{t}}(r)=e^{t(e^r -1)}E[e^{(e^{r}-1)e^{Y_t}}].
\end{eqnarray}
By applying  (\ref{Nt1}) to (\ref{St}), we have
$$
M_{S_t}(r)=e^{t(M_x(r) -1)}E[e^{(e^{lnM_{x}(r)}-1)e^{Y_t}}]=e^{t(M_x(r) -1)}E[e^{(M_{x}(r)-1)\Lambda_t}].
$$
where\ $M_{x}(r)=\frac{1}{1-r\theta}$, $\Lambda_t=$\  $e^{Y_t}$obeys log normal distribution. The characteristic function of the log normal distribution is

\begin{eqnarray}\label{duishuzhengtai}
\varphi(t)=\Sigma^{\infty}_{n=0}\frac{(it)^n}{n!}e^{n\mu^,+\frac{n^{2}\sigma^{,2}}{2}}.
\end{eqnarray}
Hence, we have the moment generating function of \ $S_t$,
\begin{eqnarray}\label{2}
M_{S_t}(r)=e^{t(M_x(r) -1)}\Sigma^{\infty}_{n=0}\frac{(e^{ln\frac{1}{1-r\theta}}-1)^n}{n!}e^{n\mu^,+\frac{n^{2}\sigma^{,2}}{2}}
=e^{t(M_x(r) -1)}\Sigma^{\infty}_{n=0}\frac{(\frac{1}{1-r\theta}-1)^n}{n!}e^{\frac{n\mu}{1-\alpha}+\frac{n^{2}\sigma^2}{2(1-\alpha^2)}}.\nonumber
\end{eqnarray}
From this we get the third-order origin moment of\ $S_{t}$,
\begin{equation}\label{E3}
E(S_{t}^3)=M^{(3)}_{S_t}(r)=\theta^3(6t+6t^2 +t^3) +3\theta^3(2t+t^2)e^{\frac{\mu}{1-\alpha}+\frac{\sigma^2}{2(1-\alpha^2)}}+3\theta^3 t e^{\frac{2\mu}{1-\alpha}+\frac{4\sigma^2}{1-\alpha^2}}+\theta^3 e^{\frac{3\mu}{1-\alpha}+\frac{9\sigma^2}{2(1-\alpha^2)}}.
\end{equation}

\subsection{\emph{Parameter Estimation}}

We figure out the parameters of the AR(1) series by the method of moment estimation. Through the observed data of the total claims which are \ $S_1,\ S_2, \cdot \cdot \cdot ,\ S_{n}$, the first origin moment, the second origin moment and the third order moments of the sample data are further obtained, which are recorded as\ $a_1,\ a_2,\ a_3$, $a_j =\frac{1}{n}\Sigma_{i=1}^{n}S_i^j,\ j=1,\ 2,\ 3$ . Let

\begin{eqnarray}\label{1}     
\left\{                        
\begin{array}{lll}       
a_1=E(S_t), \\  
a_2=E(S_t^2), \\
a_3=E(S_{t}^3). \\

\end{array}              
\right.                       
\end{eqnarray}
Therefore, we can get the estimation the parameters of\ $\alpha,\ \mu\ and\ \sigma^2$\ .

\section{Adjustment Coefficient and Ruin Probability of the Dependent Model }

In the classical model, the calculation of ruin probability is often difficult, therefore we calculate the upper bound of the ruin probability.\ This upper bound often depends on the adjustment coefficient. First we calculate the adjustment  coefficient of the surplus model.
Since

$$
U(t)=u+ct-S_{t},
$$where\ $U(t)$\ represents surplus process,\ c\ represents premium income of the unit time,\ u\ is the initial surplus. Because the $\{S_t,\ t=1,\ 2\cdot\cdot\cdot\}$  is an independent increment process, $\{U(t),\ t=1,2\cdot\cdot\cdot\}$ is an independent increment process.
Let $\mathscr{F}_t=\sigma\{U_s;s\le t\}$, we have\begin{eqnarray}\label{-rUt}
E[e^{-rU(t+1)}|\mathscr{F}_t]&=&E[e^{-r\{u+c(t+1)-S_{t+1}\}}]=e^{-ru}e^{-r\cdot c\cdot (t+1)}E(e^{rS_{t+1}})\nonumber\\
&=&e^{-ru} E[e^{\Lambda_t(M_x(r) -1)}]e^{-r\cdot c\cdot (t+1)}e^{(t+1)[M_x(r) -1]},
\end{eqnarray}
%
Based on the thought of the definition of R, in order to make the \ $\{e^{-rU (T)},\ t>0 \}$\ a martingale, there must be $E[e^{-rU(t+1)}|\mathscr{F}_t]=e^{-rU_t}$. That is $E[e^{-rU(t+1)}]=E[e^{-rU_t}]$. By (\ref{-rUt}), we need to have $e^{-rct} \cdot e^{t[M_x (r)-1]} =1,\forall t >0 $.
Therefore, we define the positive solution of
\begin{equation}\label{Rfangcheng}
e^{t[M_x (r)-1]}=e^{rct} \nonumber
\end{equation}
 is the adjustment coefficient.


It is easy to know that the adjustment coefficient \ R\ is the solution of  equation \ $rc= M_x (r)-1=\frac{1}{1-r\theta} -1$,
hence we get\ R,
$$
R=\frac{c- \theta}{c\cdot \theta},
$$

In the current insurance industry, c always needs to satisfy the condition which is
$$
c>\Lambda_t EX_{i}.
$$
It is easy to know that c needs to satisfy
$E(c)>E[\Lambda_t EX_{i}]=E(X_{i})E (\Lambda_t)$, namely

$$
c>e^{\frac{\mu}{(1-\alpha)}+\frac{\sigma^2}{2(1-\alpha^2)}}\theta.
$$
Thus $c>\theta$.


According to the upper section of the adjustment factor, we consider the calculation of ruin probability. We give the conclusion of ruin probability by Thm\ref{thm3.1}.

\begin{thm}\label{thm3.1}Consider the risk model( see (\ref{moxing})) with hypothesis $N_{t}  \sim Pois(\Lambda_{t}+t) $, the exponential upper bound of ruin probability is \ $\psi(u) \le e^{-Ru}E[e^{\Lambda_t [M_x (R)-1]}]$.
\end{thm}

{\textbf{Proof}}: Note that  \begin{equation}\label{pochangailv}
E[e^{-rU(t)}]=E[e^{-rU(t)}|T\le t]P\{T\le t\}+E[e^{-rU(t)}|T>t]P\{T> t\}\ >\ E[e^{-rU(t)}|T\le t]P\{T\le t\},
\end{equation} where T represents the time of  ruin. Since 
$$
-rU(t)=-ru-rct+rS_t,
$$
we have\begin{eqnarray}\label{-rU(t)}
E[e^{-rU(t)}]&=&E[e^{-r\{u+ct-S_{t}\}}]=e^{-ru}e^{-r\cdot c\cdot t}E(e^{rS_{t}})\nonumber\\
&=&e^{-ru} E[e^{\Lambda_t(M_x(r) -1)}]e^{-r\cdot c\cdot t}e^{t[M_x(r) -1]}.
\end{eqnarray}
We know that \ R\ is the solution of $e^{t[M_x (r)-1]}=e^{rct}$. Then we put R into (\ref{-rU(t)}) leads to

$$
E[e^{-RU(t)}]=E[e^{-Ru-Rct+R S_t}]=e^{-Ru}E[e^{\Lambda_t(M_x(r) -1)}].
$$

First of all, notice that
\begin{eqnarray}\label{St2}
E[e^{(S_{t}-S_{T}) \cdot r}\ |\ T]&=&E[e^{\Sigma_{i=1}^{N_t-N_T} X_{i} \cdot r}\ |\ T]
=E[\Pi_{i=1}^{N_{t-T}} e^{X_i \cdot r}\ |\ T]\nonumber\\
&=&E\{[M_x (r)]^{N_{t-T}}\ |\ T\}=E[e^{N_{t-T} \cdot lnM_x (r)}\ |\ T].\nonumber
\end{eqnarray}
Then,
\begin{eqnarray}\label{Nt2}
E[e^{N_{t-T} \cdot r}| T]=E[e^{(\Lambda_t -\Lambda_{T} +t-T)(e^r -1)}| T].\nonumber
\end{eqnarray}
In this way, we get 
\begin{eqnarray}\label{2}
E[e^{(S_{t}-S_{T}) \cdot r}\ |\ T]=E[e^{(\Lambda_t -\Lambda_{T} +t-T)(M_x (r) -1)}\ |\ T].\nonumber
\end{eqnarray}
Note that,
$$
U(t)=U(T)+U(t)-U(T)=U(T)+c(t-T)-[S_t-S_T],
$$
$$
-rU(t)=-rU(T)-rc(t-T)+r[S_t-S_T].
$$
Hence,
\begin{eqnarray}
E[e^{-rU(t)}|T\le t]&=& E[E[e^{-rU(t)}\ |\ T]\ |\ T \le t] \nonumber\\
&=&E[E[e^{-rU(T)-rc(t-T)+r[S_t-S_T]}\ |\ T]\ |\ T\le t]\nonumber\\
&=&E[E[e^{-rU(T)}e^{-rc(t-T)}e^{(t-T)[M_x (r)-1]}e^{(\Lambda_t-\Lambda_T) [M_x (r)-1]}\ |\ T]\ |\ T\le t]\nonumber\\
&=&E[e^{-rU(T)}e^{-rc(t-T)}e^{(t-T)[M_x (r)-1]}e^{(\Lambda_t-\Lambda_T) [M_x (r)-1]}\ |\ T\le t].\nonumber
\end{eqnarray}
We put \ R\ which is the solution of the  \ $e^{t[M_x (r)-1]}=e^{rct}$ into the equation, it follows that,
\begin{eqnarray}\label{-rut2}
E[e^{-rU(t)}|T\le t]&=&E[e^{-RU(T)}e^{(\Lambda_t-\Lambda_T) [M_x (R)-1]}|T\le t].
\end{eqnarray}
We multiply (\ref{-rut2}) by  $P\{T\le t\}$. Letting \ t\ $\rightarrow$ $+\infty$ yeilds \ $P\{T\le t\}$ $\rightarrow$ \ $ \psi(u)$, therefore
$$
E[e^{-rU(t)}|T\le t]P\{T\le t\}=E[e^{-RU(T)}e^{(\Lambda_t-\Lambda_T) [M_x (R)-1]}|T< \infty] \psi(u).
$$
Thus (\ref{pochangailv}) simplifies to
$$
\psi(u)\le \frac{e^{-Ru}E[e^{\Lambda_t [M_x (R)-1]}]}{E[e^{-RU(T)}e^{(\Lambda_t-\Lambda_T) [M_x (R)-1]}|T< \infty]},
$$
for all\ $u\ge0$. As \ $T\le t$, $N_T \le N_t$. Hence,\ $E[N_T] \le E[N_t]$, namely\ $\Lambda_T \le \Lambda_t$.


For
$$
M_x (R)-1= \frac{1}{1-\frac{c-\theta}{c}}-1=\frac{c}{\theta}-1\ >\ 0.
$$
In this case, we have$(\Lambda_t-\Lambda_T) [M_x (R)-1]\ge 0$.

Further, when  R is the solution of equation(\ref{Rfangcheng}) and\ $T< \infty $\ ,\ $U(T)<0$. We have
${E[e^{-RU(T)}e^{(\Lambda_t-\Lambda_T) [M_x (R)-1]}|T< \infty]}\ge 1$.
Meanwhile, note that\ $\Lambda_t$\ follows model(\ref{Lambda_{t}}) and it is a stationary series, therefore \ $E[e^{\Lambda_t [M_x (R)-1]}]$\ is a constant and we can obtain it through (\ref{duishuzhengtai}). We get
\begin{equation}
\psi(u) \le e^{-Ru}E[e^{\Lambda_t [M_x (R)-1]}] 
\end{equation}
which finally leads to the assertion.   $\Box$

\section{Numerical Simulation}
To check up the accuracy of our model using moment estimation method, we use AR(1) series for example and select \ $\theta=0.5$, true-value $\alpha=0.6$ ,  $\mu=0.8$ and $\sigma^2=0.4$ as parameters. Assuming that the initial value \ $ln\Lambda_0=\frac{\mu}{1-\alpha}=2$, in the environment of MATLAB , we try to do simulation towards $\alpha$, $\sigma$ and $\mu$ by trying five turns, twenty turns and fifty turns and figure out the average value. The results of simulation are as follows in table 1.
\begin{table}[!htb]
\centering
\caption{Simulation results of moment estimation of the model parameter under different sample sizes}\label{tab:abc}
\begin{center}
\begin{tabular}{cc|ccc|ccc|ccc}
\hline
 &   &        & $n=5$      &      &     &$n=20$    &       &    & $n=50$     &      \\
para  & t-v  & estimation  & deviation   & MSE & estimation & deviation& MSE &estimation & deviation  &MSE  \\ 
\hline
$\alpha$ & 0.6  & 0.6152     & 0.0152      & 0.00082 & 0.60065    &   0.00065   &0.00046 &  0.60063  &  0.00063     & 0.0001    \\
$\mu$ & 0.8  & 0.9507      &   0.1507      & 0.02280 & 0.82460   &  0.02460     &0.00060 &  0.82280 & 0.02280     & 0.0012 \\
$\sigma^2$   & 0.4  & 0.2259  & 0.1741   & 0.03250 &  0.31230  & 0.08770  &  0.00800 & 0.40400 & 0.00400   & 0.0005 \\
\hline
\footnotetext[3]{t-v represents truth value and para represents parameter}
\end{tabular}
\end{center}
\end{table}
(where t-v represents truth value and para represents parameter)
As we can see from the table 1, the accuracy made by parameter estimation differs from each other according to sample capacity. It seems perfect towards the simulation of \ $\alpha$ and $\sigma$ , the mean square error is less than 0.001, the error upon parameter\ $\mu$\  is less than 0.01 as well, which reflects the excellent simulating effect. With the increasement of samples, the deviation of estimation becomes smaller, which demonstrates the results are consistent. Concluded, we ensure the veracity of parameter \ $\alpha$, $\sigma$\ \ $\mu$\  finally.



\newpage


\begin{thebibliography}{30}
\bibitem{Lundberg}  Lundberg F. I. Approximerad Framstallning av Sannolikhetsfunktionen. II. Aterforsakring av Kollektivrisker[M]. Uppsala: Almqvist \& Wiksells. 1903.
\bibitem{Cramer} Cramer H. On the Mathematical Theory of Risk[M]. Stockholm: Shandia Jubilee . 1930.
\bibitem{Wang D H}WANG DE-HUI, GAO JIA-XING, XU ZI-LI, et al. A Class of Ruin Probability Model with Dependent Structure [J]. Communications in Mathematical Research. 2016, 32(3):241-248.
\bibitem{Fu K A}Fu Ke-ang, QIU Yu-yang, WANG An-ding. Estimates for the ruin probability of a time-dependent renewal risk model with dependent by-claims [J]. Appl. Math. J. Chinese Univ. 2015, 30(3): 347-360.
\bibitem{5} Zhen-hua Bao, Zhong-xing Ye. Ruin probabilities in the risk proceess with random income[J].Acta Mathematicae App licatae Sinica. 2008, 24(2):195-202.
\bibitem{6} Harry Joe. Muntivariate Models and Dependence Concepts [M]. CRC Press.1997.
\bibitem{7}  GU Cong, LI Shenghong. The Ruin Probability for a Class of Dependent Risk Model[J].2011 2nd International Conference on Management Science and Engineering Advances in Artificial Intelligence, 2011,1(6):61-66.
\bibitem{8}  LI Jingbo. Based on Time-series Model of Ruin Theory[D]. Urumq: Xinjiang University, 2008.
\bibitem{9}  CHEN Jie. Ruin Probabilities of Dependent Risk Models[D]. Xiamen:Xiamen University, 2007.
\bibitem{10}  Chen Y, Ng K W. The ruin probability of the renewal model with constant interest force and interst negatively dependent heavy-tailed claims [J]. Insurance: Mathematics and Economics. 2009, 40: 415-423.
\bibitem{11}  GUO Fenglong. Research on Ruin Probabilities of Insurance Risk Models with Investment Returns and Dependence Structures[D]. Chengdu:University of Electronic Science and Technology of China, 2012.
\bibitem{12}  YU Xiaohe. Research on Financial Market Dependence Modeling and Risk Measures:An Analysis Based on the GARCH-EVT-Vine Copula Model [D]. Shanghai:East China Normal University, 2016.££Š
\bibitem{13}  MO Xiaoyun,OU Hui, ZHOU Jieming. Equivalence Theorem and Probabilistic Structure for Markov-Dependence Risk Model [J]. Journal of Quantitative Economics,2012£¬29(1): 61-64.
\bibitem{14}  ZHANG Chuanwei. The Ruin Theory of a Depenent Discrete-time Risk Moel with Subexponential Insurance Risk and Related Problems [D].Hefei: Anhui University, 2015.
\bibitem{Wang Kaiyong}  WANG Kaiyong, LIN Jinguan. Finite-time ruin probability of dependent risk modelwith constant interest rate[J]. Journal of Southeast University(Natural Science Edition),2012£¬42(6):1243-1248.
\bibitem{16} Wang Kaiyong, Wang Yuebao, Gao Qingwu.Uniform Asymptotics for the Finite-Time Ruin Probability of a Dependent Risk Model with a Constant Interest Rate[J].METHODOLOGY AND COMPUTING IN APPLIED PROBABILITY. 2013,15(1):109-124.
\bibitem{17}  Gao, Qi-bing; Wu, Yao-hua; Zhu, Chun-hua. Ruin problems in risk models with dependent rates of interest[J].STATISTICS \& PROBABILITY LETTERS. 2007, 77(8):761-768.
\bibitem{18}  Badescu.Andrei L, Cheung Eric C. K, Landriault David.DEPENDENT RISK MODELS WITH BIVARIATE PHASE-TYPE DISTRIBUTIONS[J].JOURNAL OF APPLIED PROBABILITY.  2009, 46(1):113-131.
\bibitem{19} Chen Yang, Wang Le, Wang Yuebao.Uniform asymptotics for the finite-time ruin probabilities of two kinds of nonstandard bidimensional risk models[J].JOURNAL OF MATHEMATICAL ANALYSIS AND APPLICATIONS. 2013,401(1):114-129.
\bibitem{20} Wang Yuebao, Cui Zhaolei, Wang Kaiyong. Uniform asymptotics of the finite-time ruin probability for all times[J].JOURNAL OF MATHEMATICAL ANALYSIS AND APPLICATIONS.JUN 1 2012 ,390(1):208-223.
\bibitem{21} Lefevre Claude, Loisel Stephane.Finite-Time Ruin Probabilities for Discrete, Possibly Dependent, Claim Severities[J].METHODOLOGY AND COMPUTING IN APPLIED PROBABILITY. 2009,11(3):425-441.
\bibitem{22} Gao Qi-bing, Wu, Yao-hua, Zhu Chun-hua. Ruin problems in risk models with dependent rates of interest[J].STATISTICS \& PROBABILITY LETTERS. 2007,77(8):761-768.
\bibitem{23} Fu Ke-Ang, Ng Cheuk Yin Andrew. Asymptotics for the ruin probability of a time-dependent renewal risk model with geometric Levy process investment returns and dominatedly-varying-tailed claims[J]. INSURANCE MATHEMATICS \& ECONOMICS. 2014 ,56:80-87.
\bibitem{24} Li Xiaohu,  Wu Jintang, Zhuang Jinsen. Asymptotic Multivariate Finite-time Ruin Probability with Statistically Dependent Heavy-tailed Claims[J]. METHODOLOGY AND COMPUTING IN APPLIED PROBABILITY. 2015,17(2):463-477.
\bibitem{25} Yao Dingjun, Wang Rongming. Upper bounds for ruin probabilities in two dependent risk models under rates of interest[J]. APPLIED STOCHASTIC MODELS IN BUSINESS AND INDUSTRY. 2010,26(4):362-372.
\bibitem{26}  Hansjorg Albrecher, Onno J. Boxma. A ruin model with dependence between claim sizes and claim intervals [J].Insurance: Mathematics and Economics. 2004, 35(2): 245¨C254.
\bibitem{27} Qingbin Menga,Xin Zhangb, Junyi Guo. On a risk model with dependence between claim sizes and claim intervals [J]. Statistics \& Probability Letters. 2008,78(13):1727¨C1734.
\end{thebibliography}
\end{document}